\def\eg{{\frenchspacing\it e.g.}}
\def\etc{{\frenchspacing\it etc.}}

\def\beq#1{$$}
\def\eeq{$$}

\def\fig#1{Figure~\ref{#1}}

\documentstyle[prl,aps,epsf,floats]{revtex}
\draft
\begin{document}
\twocolumn[\hsize\textwidth\columnwidth\hsize\csname@twocolumnfalse\endcsname

\title{100 Years of the Quantum}

\author{Max Tegmark}

\address{Dept. of Physics, Univ. of Pennsylvania, 
Philadelphia, PA 19104; max@physics.upenn.edu}

\author{John Archibald Wheeler}

\address{Princeton University, Department of Physics,  
Princeton, NJ 08544; jawheeler@pupgg.princeton.edu\\$\>$}

\date{An abbreviated version of this article, with much better graphics, \\
was published in the Feb.~2001 issue of {\it Scientific American}, p.68-75.}
\maketitle

\begin{abstract}
\noindent{\bf Abstract:} 
As quantum theory celebrates its 100th birthday,
spectacular successes are mixed with outstanding puzzles
and promises of new technologies.
This article reviews both 
the successes of quantum theory and the ongoing
debate about its consequences for issues ranging from quantum
computation to consciousness, parallel universes and the
nature of physical reality. 
We argue that modern experiments and 
the discovery of decoherence have have shifted prevailing quantum 
interpretations away from wave function collapse towards unitary physics,
and discuss quantum processes in the framework of a tripartite 
subject-object-environment decomposition.
We conclude with some speculations on the bigger picture 
and the search for a unified theory of quantum gravity. 
\end{abstract}
\bigskip
] 


\setcounter{secnumdepth}{0}


{\it ``...in a few years, all the great physical constants will have been
approximately estimated, and [...] the only occupation which
will then be left to the men of science
will be to carry these measurement to another place of decimals.''}
As we enter the 21st century amid much brouhaha 
about past achievements, this
sentiment may sound familiar. Yet the quote is from James Clerk
Maxwell and dates from his 1871 Cambridge inaugural
lecture expressing the mood prevalent at the time
(albeit a mood he disagreed with). 
Three decades later, on December 14, 1900, Max
Planck announced his famous formula on the blackbody spectrum, the
first shot of the quantum revolution. 

This article reviews both
the spectacular successes of quantum theory and the ongoing
debate about its consequences for issues ranging from quantum
computation to consciousness, parallel universes and the very
nature of physical reality. 

\section{The ultraviolet catastrophe}

In 1871, scientists had good reason for their optimism. Classical
mechanics and electrodynamics had powered the industrial
revolution, and it appeared as though their basic equations
could describe essentially all physical systems. Yet 
some annoying details tarnished this picture. The amount of
energy needed to heat very cool objects was smaller than
predicted and the calculated spectrum of a glowing hot object
didn't come out right. In fact, if you took the classical
calculation seriously, the prediction was the so-called
ultraviolet catastrophe: that you would get blinded by light
of ultraviolet and shorter wavelengths 
when you looked at the burner on your stove!

In his 1900 paper, Planck succeeded in deriving the correct
shape of the blackbody spectrum which now bears his name,
eliminating the ultraviolet catastrophe. However, this involved
an assumption so bizarre that even he distanced himself
from it for many years afterwards: that energy was only emitted
in certain finite chunks, or ``quanta''. Yet this strange
assumption proved extremely successful. Inspired by Planck's
quantum hypothesis, Peter Debye showed that the strange thermal
behavior of cold objects could be explained if you assumed that 
the vibrational energy in solids could only come in discrete
chunks. In 1905, Einstein took this bold idea one step further. 
Assuming that radiation could only transport energy in such
chunks, ``photons'', he was able to explain the so-called
photoelectric effect, which is related to the processes
used in present-day solar cells and the image sensors
in digital cameras.

\section{The hydrogen disaster}

In 1911, physics faced another another great embarrassment.
Ernest Rutherford had convincingly argued that atoms consisted of
electrons orbiting a positively charged nucleus much like a
miniature solar system. However, electromagnetic theory predicted
that such orbiting electrons would radiate away their energy,
spiraling inward until they got sucked into the atomic nucleus
after about a millionth of a millionth of a second. Yet hydrogen
atoms were known to be eminently stable. Indeed, this was the
worst quantitative failure so far in the history of physics,
under-predicting the lifetime of hydrogen by some forty orders of
magnitude!

Niels Bohr, who had come to Manchester to work with Rutherford,
made a breakthrough in 1913. By postulating that the amount of
angular momentum in an atom was quantized, the electrons were
confined to a discrete set of orbits, each with a definite energy.
If the electron jumped from one orbit to a lower one, the energy
difference was sent off in the form of a photon. If the electron
was in the innermost allowed orbit, there were no orbits with
less energy to jump to, so the atom was stable. In addition,
Bohr's theory successfully explained a slew of spectral lines
that had been measured for Hydrogen. It also worked for the
Helium atom, but only if it was deprived of one of its two
electrons. Back in Copenhagen, Bohr got a letter from Rutherford
telling him he had to publish his results. Bohr wrote back that
nobody would believe him unless he could explain the spectra of
all the atoms. Rutherford replied, in effect: Bohr, you explain
hydrogen and you explain helium and everyone will believe all the
rest.

Bohr was a warm and jovial man, with a talent for leadership,
and that business of explaining all the rest 
soon become the business of the group that rose 
around him at Copenhagen. 
The second author
had the privilege to work there on nuclear physics 
from September 1934 to June 1935,
and on arrival asked a workman who was trimming vines running
up a wall where he could find Bohr. ``I'm Niels Bohr'', the man replied.

\section{The equations fall into place}

Despite these early successes, physicists still didn't 
know what to make of these strange and seemingly ad hoc 
quantum rules. What did they really mean?

In 1923, Louis de Broglie proposed an answer in his Ph.D. thesis:
that electrons and other particles acted like standing waves.
Such waves, like vibrations of a guitar string, can only occur with
certain discrete (quantized) frequencies. 
The idea was so new that the
examining committee went outside its circle for advice on the acceptability
of the thesis. Einstein gave a favorable opinion and the thesis was
accepted. 
In November 1925, Erwin Schr\"odinger gave a seminar 
on de Broglie's work in Zurich. When he was finished, 
Debye said in effect, ``You speak about waves. But where is the wave equation?''
Schr\"odinger went on to produce and publish his famous wave 
equation, the master key for so much of modern physics.
An equivalent formulation involving matrices was provided by 
Max Born, Pasquale Jordan and Werner Heisenberg around the same time.
With this new powerful mathematical underpinning, quantum theory 
made explosive progress. Within a few years, a host of hitherto
unexplained measurements had been successfully explained, 
including spectra of more complicated atoms and
various numbers describing properties of chemical reactions.

\begin{figure}[tbp]
\centerline{{\vbox{\epsfxsize=8.6cm\epsfbox{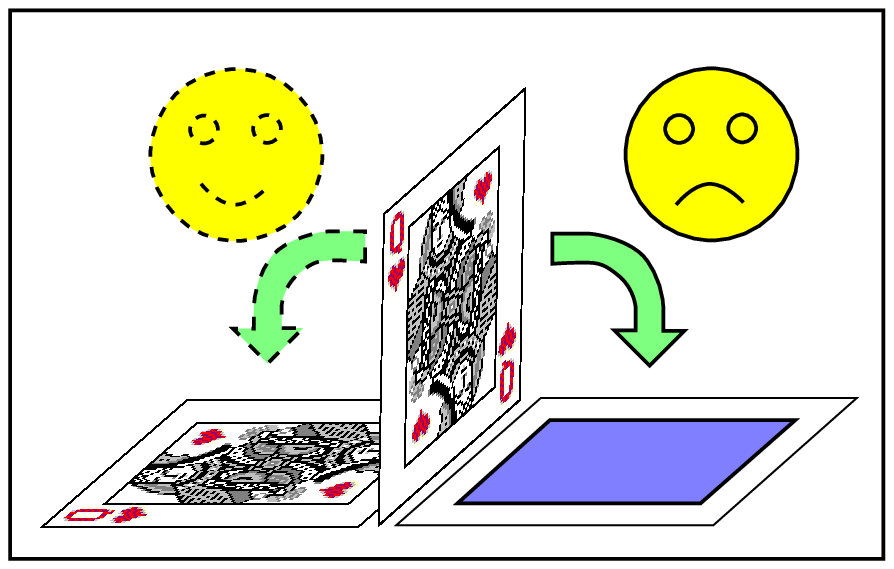}}}}
\bigskip
\centerline{{\vbox{\epsfxsize=8.6cm\epsfbox{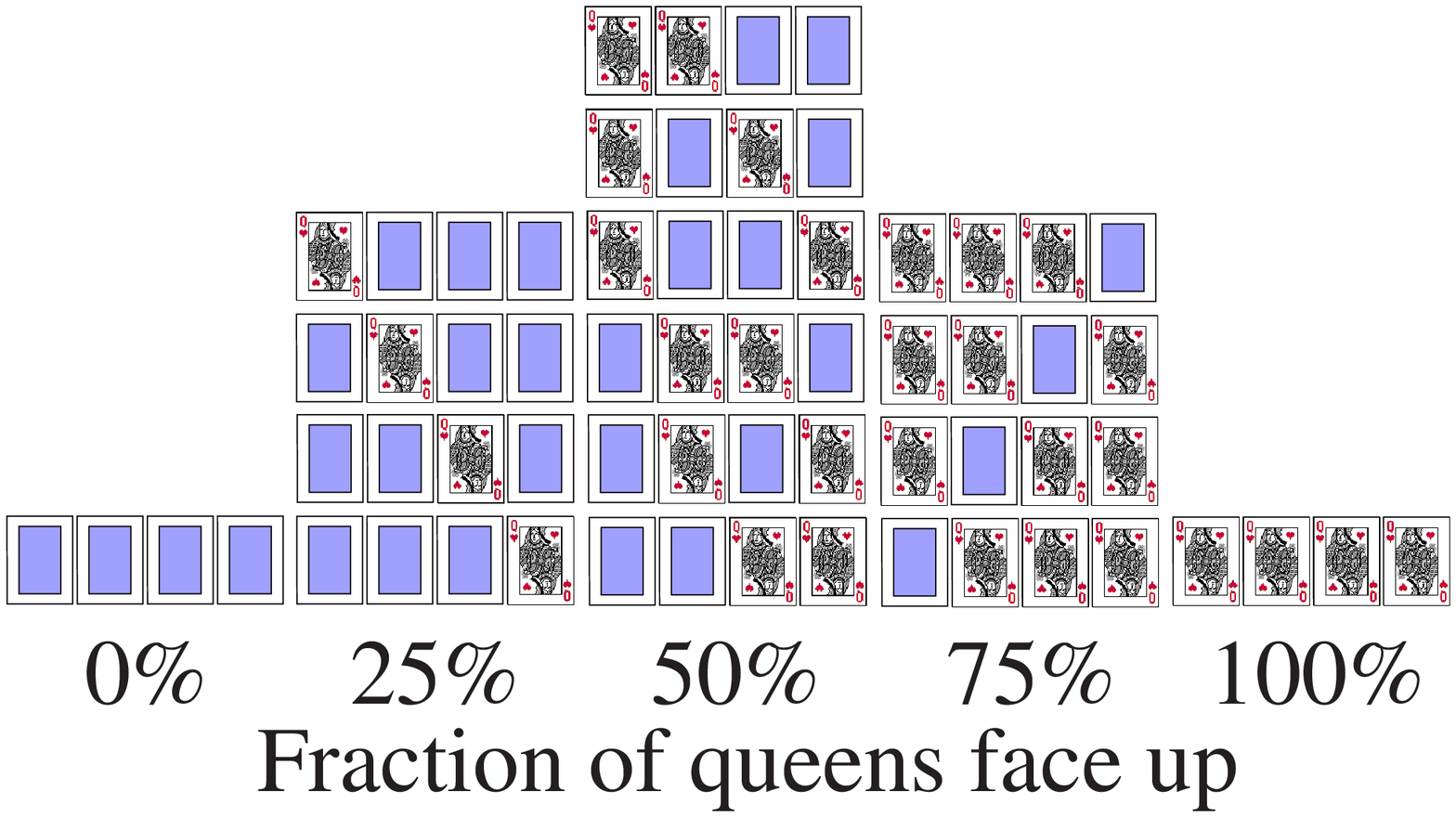}}}}
\bigskip
\caption{
According to quantum physics, a card perfectly balanced on its
edge will by symmetry fall down in both directions at once, 
in what is known as a
``superposition''. If an observer has bet money on the queen landing
face up, the state of the world will become a superposition of
two outcomes: her smiling with the queen face up and
her frowning with the queen face down. 
In each case, she is unaware of the other outcome and feels 
as if the card fell randomly.
If our observer repeats this experiment with four cards, 
there will be $2\times 2\times 2\times 2=16$ outcomes
(see figure).
In almost all of these cases, it will appear to her 
that queens occur randomly, with about 50\% probability.
Only in 2 of the 16 cases will she get the same result all four times.
According to a 1909 theorem by the French mathematician Borel,
she will observe queens 50\% of the time in almost 
all cases (in all cases except for what mathematicians
call a set of measure zero)
in the limit where she repeats the card experiment 
infinitely many times. Almost all of the observers in the
final superposition will therefore conclude that the laws of 
probability apply even though the underlying physics is not random
and God does not play dice.
}
\label{CardFig}
\end{figure}


But what did it all mean?
What was this quantity, the ``wave function'', 
which Schr\"odinger's equation described?
This central puzzle of quantum mechanics
remains a potent and controversial issue to this day. 

%
%
%

Max Born had the dramatic insight that the wave function should be 
interpreted in terms of probabilities. If we measure the location of 
an electron, the probability of finding it in a given region
depends on the intensity of its wave function there. This 
interpretation suggested that
a fundamental randomness was
built into the laws of nature. Einstein was deeply unhappy with
this interpretation, and expressed his preference for a 
deterministic Universe with the oft-quoted remark
``I can't believe that God plays dice''.

\section{Curious cats and quantum cards}

Schr\"odinger was also uneasy.
Wave functions could describe combinations of different states, 
so-called superpositions. 
For example, an electron could be in a superposition of several different 
locations. Schr\"odinger pointed out that if 
microscopic objects like atoms could be in strange superpositions, so could
macroscopic objects, since they are made of atoms.
In particular, seemingly innocent ``microsuperpositions'' could 
turn into ``macrosuperpositions''.
%
As a baroque example, he described the famous thought experiment
where a nasty contraption kills a cat if a radioactive atom decays.
Since the radioactive atom eventually enters a superposition
of decayed and not decayed, it produces a cat which is both dead and alive 
in superposition.

Figure 1 shows a simpler version of this {\it Gedanken} 
experiment that we will call
Quantum Cards, again turning a microsuperposition into a macrosuperposition. 
You simply take a card 
with a perfectly sharp bottom edge and balance it on its edge on a table.
According to classical physics, it will in principle stay balanced forever.
(In practice, this unstable card will of course get toppled in no time
by say a tiny air current, so you could take a card with a thick bottom edge 
and use Schr\"odinger's radioactive atom trigger 
to nudge it one way or the other.)
According to the Schr\"odinger equation, it will fall down in a few seconds
even if you do the best possible job of balancing it, because 
the Heisenberg uncertainty principle states that it cannot be in
only one position (straight up) without moving.
Yet since the initial state was left-right symmetric, the final
state must be so as well. 
The implication is that it falls 
down in both directions at once, in superposition.
If you could perform this thought experiment, 
you would undoubtedly find that classical
physics was wrong and the card fell down. But you would always see 
it fall down to the left {\it or} to the right, seemingly at random,
never to the left and to the right simultaneously as the Schr\"odinger 
equation might have you believe. This apparent contradiction goes to the
very heart of one of the original and most enduring mysteries of
quantum mechanics.


The Copenhagen Interpretation of quantum mechanics, which
evolved from discussions 
between Bohr and Heisenberg in the late 1920s, addresses the mystery
by asserting that observations, or measurements, are special.
So long as the balanced card is unobserved, its wave function evolves by 
obeying the Schr\"odinger equation -- a continuous and smooth evolution 
that is called ``unitary'' in mathematics and has several very attractive
properties. Unitary evolution produces the superposition where the card 
has fallen down both to the left and to the right. The act of observing 
the card, however, triggers an abrupt change in its wave function, 
commonly called a ``collapse'': the observer sees the card in one 
definite classical state (face up or face down) and from then onward 
only that portion of the wave function survives.
Nature supposedly decided which particular state to collapse into
at random, with the probabilities 
determined by the wave function.

Although this provided a strikingly successful calculational recipe, 
there was a lingering feeling that there ought to be some 
equation describing when and how this collapse occurred.
Many physicists took this to mean that there is something fundamentally
wrong with quantum mechanics, and that it would soon be replaced by 
some even more fundamental theory that provided such an equation.
So rather than dwell on ontological
implications of the equations, most workers forged ahead to work out
their many exciting applications and to tackle pressing unsolved 
problems of nuclear physics.

That pragmatic approach proved stunningly successful. Quantum mechanics was 
instrumental in predicting antimatter, understanding radioactivity 
(leading to nuclear power), accounting for materials such as semiconductors,
explaining superconductivity, 
and describing interactions such as those between light and matter
(leading to the invention of the laser) and of radio waves and nuclei
(leading to magnetic resonance imaging).
Many successes of quantum mechanics involve its extension, 
quantum field theory, which forms the foundation of elementary particle
physics all the way to the present-day experimental 
frontiers of neutrino oscillations and the search for 
the Higgs particle and supersymmetry.

\section{Many worlds or many words?}

By the 1950's, this ongoing parade of successes had 
made it abundantly clear that quantum theory 
was far more than a short-lived temporary fix.
And so, in the mid 1950's, 
a Princeton graduate student 
named Hugh Everett III decided to revisit the collapse postulate
in his Ph.D. thesis.
Everett pushed the quantum idea to its extreme by asking
the following question: 
{\it ``What if the time-evolution of the entire Universe is always unitary?''}
After all, if quantum mechanics suffices to describe the Universe, 
then the present state of the Universe is described by a wave function
(an extraordinarily complicated one).
In Everett's scenario,
that wave function would always evolve in a deterministic
way, leaving no 
room for wave function collapse or God playing dice.

Instead of getting collapsed by measurements, 
seemingly innocent microscopic superpositions
would rapidly get amplified into most 
Byzantine macroscopic superpositions. Our quantum card in 
Figure 1 would really be in two places at once. 
Moreover, a person looking at the card would enter a 
superposition of two different mental states, each 
perceiving one of the two outcomes! 
If you had bet money on the queen coming face up, 
you would end up in a superposition of smiling and frowning.
Everett's brilliant insight was that the observers in such
a crazy deterministic but schizophrenic quantum world could 
perceive the plain old reality that we are familiar with,
as described in \fig{CardFig}.
Most importantly, they would perceive an apparent randomness
obeying precisely the right probability rules, as the bottom panel
of \fig{CardFig} illustrates. 
The situation is more complicated, and still controversial, for the 
asymmetric case when the probabilities for different outcomes are not equal. 

Everett's viewpoint became known as the 
``many worlds'' or, perhaps more appropriately, 
``many minds'' interpretation of quantum mechanics
because 
each of one's superposed mental states perceives its own world.
This viewpoint simplifies the underlying theory by removing 
the collapse postulate, implying that there 
is no new undiscovered physics that makes these superpositions go away. 
The price it pays for this theoretical simplicity is the conclusion that 
these parallel perceptions of reality
are all equally real, so in a sense it involves
less words at the expense of more worlds.

\section{Experimental verdict:\\the world {\it is} weird}

Everett's work was largely disregarded for about two decades.
The main objection was that it was too weird, demoting
to mere approximations
the familiar classical concepts upon which
the Copenhagen 
interpretation was founded. 
Many physicists hoped that a deeper theory would be discovered,
showing that the world was in some sense classical after all,
free from oddities like large objects being in two places at once.
However, such hopes were largely shattered by a series of 
new experiments.

Could the apparent quantum randomness be replaced by some kind
of unknown quantity carried about inside particles, so-called
``hidden variables''? CERN theorist John Bell showed that in this case, 
quantities that could be measured in certain difficult experiments 
would inevitably disagree with the standard quantum predictions.
After many years, technology allowed researchers to conduct these experiments
and eliminate hidden variables as a possibility.

\begin{figure}[pbt]
\centerline{{\vbox{\epsfxsize=9.0cm\epsfbox{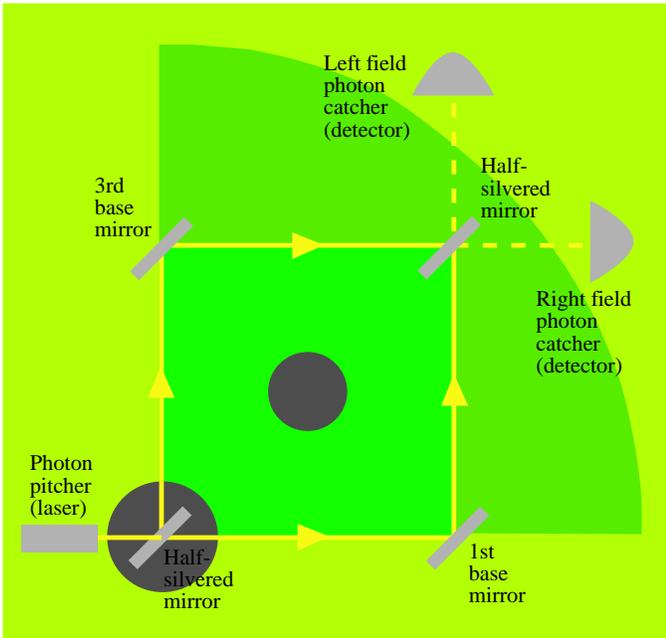}}}}
\caption{
The delayed choice experiment.
The figure shows a so-called Mach-Zender interferometer in
the guise of a baseball diamond. The half-silvered mirror
on home plate reflects half the light that strikes it towards 3rd base
and lets the rest through towards 1st base. Two ordinary mirrors
then reflect this light towards 2nd base, where the two beams
strike another half-silvered mirror. The setup is such that 
the two beams headed for left field will cancel each other through 
destructive interference, i.e., all photons fired from 
home plate will be detected in right field, none in left field.
This implies that each photon took {\it both} the 1st and 3rd base
routes in superposition.
If we remove the half-silvered mirror at second base, we detect half of the
photons in left field and half in right field, and we know which
baselines each photon traveled along.
We can therefore choose whether the individual 
photons should act schizophrenically
or not. Indeed, we can delay this choice until seemingly 
after the fact!
Let us turn on the light source for only a billionth of a 
second, during which time it emits, say, 1000 photons. 
At the speed of light, the photons travel about a foot during this time.
Let us therefore wait a leisurely 10 or 20 billionths of 
a second after the light source is turned off before we decide 
which experiment we want to do. The two photon convoys headed
for 1st and 3rd base will be separated by many meters by then,
unable to communicate with each other.
If we want to demonstrate that each photon followed both routes at once,
we need only wait and note that no photons make it to left field.
If we want to find out which way each photon went, we swiftly remove
the 2nd base mirror before the photons have had time to reach it.
}
\label{BaseballFig}
\end{figure}

A ``delayed choice'' experiment proposed by the second author in 1978
(see \fig{BaseballFig}) was successfully carried out by 
Carroll Alley, 
Oleg Jakubowics and William Wickes 
in 1984, showing that not only can a photon be in two places
at once, but we can decide whether it should act schizophrenically
or classically seemingly after the fact!

The simple double slit interference experiment,
hailed by Feynman as the mother of all quantum effects, was
successfully repeated for ever larger objects: atoms, small
molecules and most recently a carbon-60 ``Buckey Ball''.
After this last feat, Anton Zeilinger's group in Vienna 
has even started discussing doing it with a virus.
If we imagine, as a {\it Gedanken} experiment, that 
this virus has some primitive kind of 
consciousness, then
the many worlds/many minds interpretation seems unavoidable,
as has been emphasized by Dieter Zeh.
An extrapolation to superpositions involving other sentient beings 
such as humans would then be merely a quantitative rather 
than a qualitative one.


In short, the experimental verdict is in: the weirdness of the quantum
world is real, whether we like it or not.
There are in fact good reasons to like it: 
this very weirdness may offer useful
new technologies. According to a recent estimate, 
about 30\% of the U.S. gross national product is 
now based on inventions made possible by quantum mechanics. 
Moreover, if physics really is unitary (if the wave function
never collapses), quantum computers can in principle 
be built that take advantage of such superpositions to
make certain calculations much faster than
conventional algorithms would allow.
For example, Peter Shor and Lov Grover have
shown that one could factor large numbers and search 
long lists faster this way.
Such machines would be the ultimate parallel computers,
in a sense running many calculations in superposition.
As David Deutsch has emphasized, 
it will be hard to deny the reality of all these parallel
states if such computers are actually built.

%
%
%
%

\section{Quantum censorship: decoherence}
\label{DecoherenceSec}

The above-mentioned experimental progress of the last few
decades was paralleled by a new breakthrough in theoretical understanding.
Everett's work had left two crucial question unanswered: first of all,
if the world actually contains bizarre macrosuperpositions, then
why don't we perceive them? 

The answer came in 1970 with a seminal paper by
Dieter Zeh of the University of Heidelberg, 
who showed that the Schr\"odinger equation itself
gives rise to a type of censorship.
This effect became known as {\it decoherence}, because
an ideal pristine superposition is said to be coherent. 
Decoherence
was worked out in great detail by Los Alamos scientist 
Wojciech Zurek, Zeh and others
over the following decades. They found
that coherent quantum superpositions persist only as long as they remain
secret from the rest of the world. Our fallen quantum card in Figure 1 is
constantly bumped by snooping air molecules, photons, \etc, which thereby
find out whether it has fallen to the left or to the right, destroying
(``decohering'') the superposition and making it unobservable. This is somewhat
analogous to the way in which interference effects in classical optics would
get destroyed if perturbations along the light path messed up the phases.

The most convenient way to 
understand decoherence is by looking at a generalization of 
the wave function called the {\it density matrix}. For every wave function, there is a corresponding 
density matrix, and there is a corresponding Schr\"odinger's 
equation for density matrices. For example, the density matrix of the 
quantum card fallen down in a superposition would look like this:
\beq{MatrixEq1}
\hbox{density matrix} = 
\left(
\begin{tabular}{cc}
$a$&$c$\\
$c^*$&$b$\\
\end{tabular}
\right).
\eeq
The numbers $a$ and $b$ are the probabilities of finding the card face up 
or face down respectively, and would both equal one half
in our case. Indeed, a density matrix having the form
\beq{MatrixEq2}
\hbox{density matrix} = 
\left(
\begin{tabular}{cc}
$a$&$0$\\
$0$&$b$\\
\end{tabular}
\right)
\eeq
would represent a familiar classical situation --- a card that had 
fallen one way or the other, but we didn't know which.
The off-diagonal numbers in the matrix ($c$ in our 
simple example) thus represent the difference between the quantum 
uncertainty of superpositions and the classical uncertainty 
(mere ignorance).
 
A remarkable achievement of decoherence theory is to explain how 
interactions between an object and its environment push the 
off-diagonal numbers essentially to zero, for all practical
purposes replacing the quantum superposition by pure classical ignorance.

If your friend observed the
card without telling you the outcome, she would according to the
Copenhagen interpretation 
collapse the superposition into classical ignorance (on your part),
replacing $c$ by zero.
Loosely speaking, decoherence calculations show that you don't need a human
observer to get this effect --- even an air molecule will suffice.

Decoherence explains why we do not routinely see quantum superpositions
in the world around us. It is not because quantum mechanics intrinsically
stops working for objects larger than some magic size.
Instead, macroscopic objects such as cats and cards are 
almost impossible to keep isolated to the extent needed to 
prevent decoherence. Microscopic objects, in contrast, are 
more easily isolated from their surroundings so that they
retain their quantum secrets and quantum behavior.

The second unanswered question in the Everett picture 
was more subtle but equally important:
what physical mechanism picks out the 
classical states  --- face up and face down for the card --- as special?
The problem was that from a mathematical point of view, 
quantum states like ``face up plus
face down'' (let's call this ``state alpha'') or ``face up minus face down'' (``state
beta'', say) are just as valid as the classical states ``face up'' or ``face down''. So just
as our fallen card in state alpha can collapse into the face up or face down states, a card
that is definitely face up --- which equals (alpha + beta)/2 --- should be able to collapse
back into the alpha or beta states, or any of an infinity of other states into which ``face
up'' can be decomposed. Why don't we see this happen? 

Decoherence answered this question as well.
The calculations showed that classical states could by defined and identified 
as simply those states that were most robust against decoherence.
In other words, decoherence does more than just make 
off-diagonal matrix elements go away.
If fact, if the alpha and beta states of our card were taken as the fundamental basis,
the density matrix for our fallen card would be diagonal to start with,
of the simple form
\beq{MatrixEq3}
\hbox{density matrix} = 
\left(
\begin{tabular}{cc}
$1$&$0$\\
$0$&$0$\\
\end{tabular}
\right),
\eeq
since the card is definitely in state alpha.
However, decoherence would almost instantaneously change the state to 
\beq{MatrixEq4}
\hbox{density matrix} = 
\left(
\begin{tabular}{cc}
$1/2$&$0$\\
$0$&$1/2$\\
\end{tabular}
\right),
\eeq
so if we could measure whether the card was in the alpha or beta-states,
we would get a random outcome.
In contrast, if we put the card in the state ``face up'', it would stay
``face up'' in spite of decoherence.
Decoherence therefore provides what Zurek has termed a 
``predictability sieve'', selecting out those states that display
some permanence and in terms of which physics has predictive power.


\section{Shifting views}

The discovery of decoherence, combined with the ever more elaborate 
experimental demonstrations of quantum weirdness, has
caused a noticeable shift in the views of physicists.
The main motivations for introducing the notions of 
randomness and wave function collapse in the first place had
been to explain why we perceived probabilities and not strange 
macrosuperpositions. After Everett had shown that 
things would appear random anyway and decoherence 
had been found to explain why we never perceived anything strange,
much of this motivation was gone. 
Moreover, it was embarrassing that nobody 
had managed to provide a testable deterministic 
equation specifying precisely when 
this mysterious collapse was supposed to occur.
Even though the wave function technically never collapses 
in the Everett view, it is generally agreed that
decoherence produces an effect that 
looks like a collapse and smells like a collapse.

An informal poll taken at a conference on quantum computation
at the Isaac Newton Institute in Cambridge in July 1999 
gave the following results:
\begin{enumerate}
\item {\it Do you believe that new physics violating the 
Schr\"odinger equation will make large quantum computers impossible?}
1 yes, 71 no, 24 undecided

\item {\it Do you believe that all isolated systems 
obey the Schr\"odinger equation
(evolve unitarily)?}
59 yes, 6 no, 31 undecided 

\item {\it Which interpretation of quantum mechanics is closest
to your own?}
\begin{enumerate}
\item
Copenhagen or consistent histories (including 
postulate of explicit collapse): 4
\item 
Modified dynamics (Schr\"odinger equation modified to 
give explicit collapse): 4
\item Many worlds/consistent histories (no collapse): 30
\item Bohm (an ontological interpretation where an auxiliary ``pilot wave''
allows particles to have well-defined positions and velocities): 2
\item None of the above/undecided: 50
\end{enumerate}
\end{enumerate}
The reader is warned of
rampant linguistic confusion in this area.
It is not uncommon that two physicists who say that
they subscribe to the Copenhagen interpretation 
find themselves disagreeing about what they mean by this.
Similarly, some view the ``consistent histories'' interpretation
(in which the fundamental objects are consistent sets of classical histories)
as a fundamentally random theory where God plays dice (as in the
recent {\it Physics Today} article by Omn\`es \& Griffith),
whereas others view it more as a way of identifying what is 
classical within the deterministic 
``many worlds'' context. Such issues undoubtedly
contributed to the large ``undecided'' vote on the last question.

This said, the poll clearly suggests that it is 
time to update the quantum textbooks:
although these infallibly list explicit non-unitary collapse as a fundamental 
postulate in one of the early chapters, the poll indicates that 
many physicists --- at least in the burgeoning field of quantum 
computation --- no longer take this seriously.
The notion of collapse will undoubtedly retain great 
utility as a calculational recipe, but an added caveat 
clarifying that it is probably an not a fundamental process 
violating the Schr\"odinger equation could save astute students
many hours of frustrated confusion.

The Austrian animal behaviorist Konrad Lorenz
mused that important scientific discoveries go though three
phases: first they are completely ignored, 
then they are violently attacked, and finally they are
brushed aside as well-known.
Although more quantitative experimental study of decoherence is clearly 
needed, it is safe to say that decoherence has now reached 
the third phase among quantum physicists --- indeed,
a large part of current quantum computing research is 
about finding ways to minimize decoherence.
The poll suggests that after spending the sixties in
phase 1, Everett's idea that physics is unitary 
(that there is no wave function collapse) is 
now shifting from phase 2 to phase 3, replacing the collapse 
interpretation as the dominant paradigm.


\section{How does it fit together?}
\begin{figure}[pbt]
\vskip-1.5cm
\centerline{{\vbox{\epsfxsize=8.5cm\epsfbox{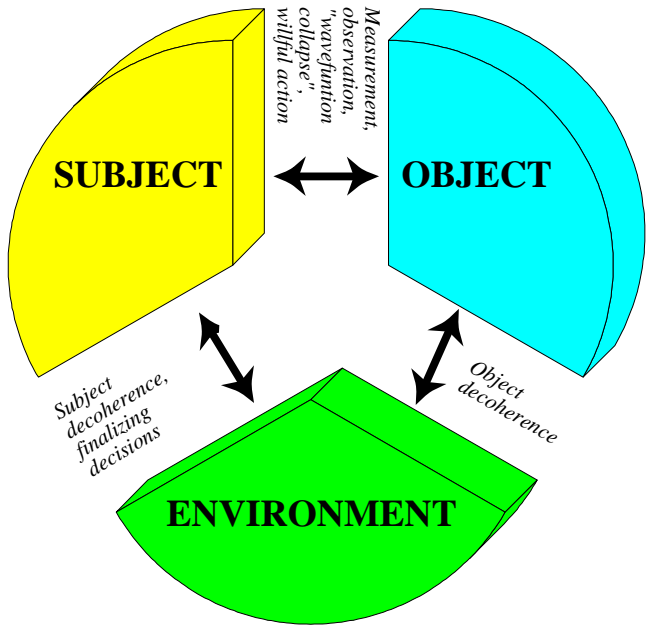}}}}
\bigskip
\caption{
An observer conveniently
decomposes the world into three subsystems:
the degrees of freedom corresponding to her subjective
perceptions (the subject),
the degrees of freedom being studied (the obj                                      ect), 
and everything else (the environment).
As indicated, the interactions between these three
subsystems can cause qualitatively very different effects,
which is why it is often useful to study them separately.
}
\label{TrinityFig}
\end{figure}

If unitarity and decoherence are taken seriously, then
it is instructive to split the Universe 
into three parts as illustrated in 
\fig{TrinityFig}.
As emphasized by Feynman,
quantum statistical mechanics splits the Universe 
(or, in physics jargon, its ``degrees of freedom'') 
into two subsystems:
the object under consideration and everything else
(referred to as the {\it environment}).
To understand processes such as measurement, we 
need to include a third subsystem as well: the {\it subject},
the mental state of the observer.
A useful standard technique is to split 
the Schr\"odinger equation that governs the 
time evolution of the Universe as a whole into
terms that describe the internal dynamics of each of
these subsystems and terms that describe interactions
between them. These different terms have qualitatively 
very different effects.

The term giving the object dynamics is normally the most important
one, so to figure out what the object will do, all the other
terms can usually be ignored. 
Consider the quantum card example in \fig{CardFig}, with the
``object'' being the (position of) the card.
In this case, the object dynamics is such that the card will
fall left and right in superposition.
When our observer looks at the card, this 
subject-object interaction will make her mental state
enter a superposition of joy and disappointment over
winning/losing her bet. However, she can never be 
aware of her schizophrenic state of mind, since
interactions between the object and the environment (in this
case air molecules and photons bouncing off of the card)
cause rapid decoherence that makes this superposition
completely unobservable.\footnote{Although the processes
of measurement and decoherence may appear different,
there is a symmetry between the 
object-subject and object-environment interactions,
involving the lack of information about
the object  (entropy): loosely speaking, the entropy of an object 
decreases while you look at it and 
increases while you don't. Decoherence is essentially
a measurement that you don't know the outcome of. 
}
It would be virtually impossible for her to
eliminate this decoherence in practice since the
card is so large, but even if she could
(say by repeating the experiment in a dark cold room with no air),
it wouldn't make any difference: 
at least one neuron in her optical nerves would enter a superposition
of firing and not firing while she looked at the 
card, and this superposition would decohere in
about $10^{-20}$ seconds according to recent calculations.

There could still be trouble, since thought processes 
(the internal dynamics of the subject system) could create
superpositions of mental states that we do not in fact 
perceive. Indeed, Roger Penrose and others have suggested that
such effects could let our brains act as 
quantum computers.
However, the fact that neurons decohere much faster than they
can process information (it takes them about 
$10^{-3}$ seconds to fire) means that if 
the complex neuron firing patterns in our brains have
anything to do with consciousness, then decoherence
in the brain will prevent us from perceiving 
weird superpositions.

As mentioned above,
we perceive only those aspects of the world that are most 
robust against decoherence. Decoherence therefore 
selects what Zurek has termed a ``pointer basis'', basically 
the familiar quantities of classical physics, as special.
Since all our observations are transmitted through neurons
from our sensory organs, the fact that
neurons decohere so fast makes them the ultimate pointer
basis. As Zeh has stressed, this justifies 
using the textbook wave function collapse postulate as a useful
``shut-up-and-calculate'' recipe:
compute probabilities as if the wave function collapses 
when we observe the object.
Strictly speaking, we constantly keep entering into superpositions of 
different mental states, but decoherence prevents us 
from noticing this --- subjectively,
we (all superposed versions of us) 
just perceive this as the slight randomness that disturbed
Einstein so much.

A basic question of course remains:
can quantum mechanics be understood in terms 
of some deeper underlying principle?
How come the quantum?

\section{Looking ahead}

\begin{figure}[pbt]
\hglue-1.7cm\centerline{{\vbox{\epsfxsize=12.0cm\epsfbox{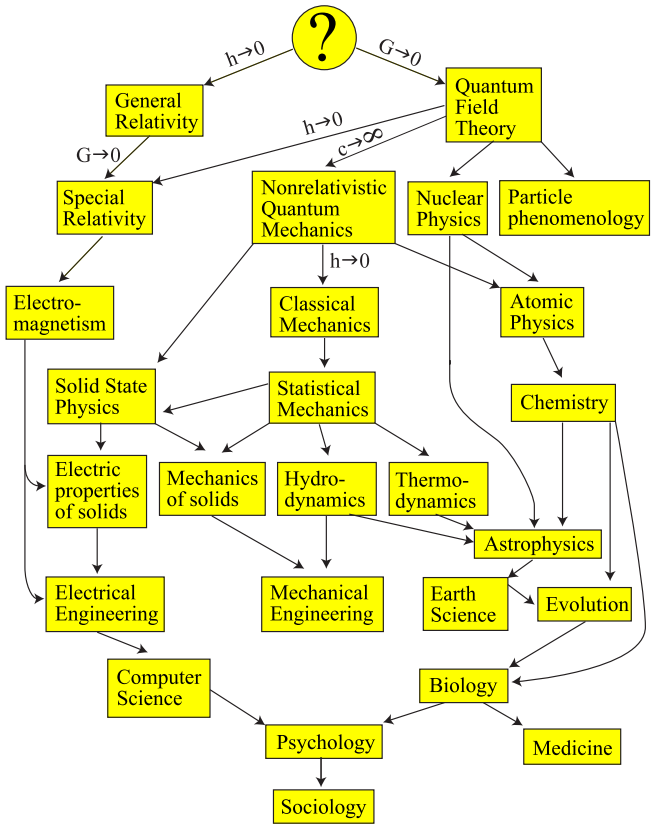}}}}
\smallskip
\caption{
Theories can be crudely organized into a family tree where
each might, at least in principle, be derivable
from more fundamental ones above it.
For example, classical mechanics can be obtained from 
special relativity in the approximation that the speed of light 
$c$ is infinite, and hydrodynamics with its concepts such as 
density and pressure can be derived from statistical 
mechanics. However, these cases where the arrows are well understood
form a minority. Although chemistry in principle should be derivable
from quantum mechanics, the properties of 
some molecules are so complicated to compute 
in practice that a more empirical 
approach is taken. Deriving biology from chemistry 
or psychology from biology would be even
more hopeless in practice.
}
\label{TreeFig}
\end{figure}

After 100 years of the quantum, let us take a step back
and make some general remarks about what may lie ahead.
Although basic issues about ontology and the ultimate nature of 
reality often crop up in discussions about how to interpret quantum
mechanics, this is probably just a piece in a larger puzzle.
As illustrated in \fig{TreeFig}, 
theories can be crudely organized in a family tree where
each might, at least in principle, be derivable
from more fundamental ones above it.

All these theories have two components: mathematical equations
and words that explain how they are connected to 
what we observe. 
Quantum mechanics as usually presented in textbooks has both 
components: some equations as well as three fundamental postulates
written out in plain English.
At each level in the hierarchy of theories, new concepts 
(\eg, protons, atoms, cells, organisms, cultures) are introduced 
because they are convenient, capturing the essence of what
is going on without recourse to the more fundamental theory
above it.  
It is important to remember, however, 
that it is we humans who introduce these concepts and the words
for them: in principle, everything could have been derived
from the fundamental theory at the top of the tree, although 
such an extreme reductionist approach would of course be
useless in practice.
Crudely speaking, the ratio of equations to
words decreases as we move down the tree, dropping near zero for 
highly applied fields such as medicine and sociology.
In contrast, theories near the top are highly mathematical, and 
physicists are still struggling to understand the concepts, if any, 
in terms of which we can understand them.

The Holy Grail of physics is to find what is jocularly
referred to as a ``Theory of Everything'', or TOE, from which
all else can be derived. If such a theory exists at all, 
it should replace the big question mark at the top of the 
theory tree. Everybody knows that something is missing
here, since we lack a consistent theory unifying gravity
with quantum mechanics.
To avoid the problem of infinite regress, where 
each set of concepts is explained in terms of more fundamental 
ones that in turn must be explained, a TOE would probably have to 
contain no concepts at all. In other words, it would have to
be a purely mathematical theory, with no explanations or
``postulates'' as in quantum textbooks
(recall that mathematicians are perfectly capable 
of --- and often pride themselves of --- studying 
abstract mathematical structures that lack any
intrinsic meaning or connection with physical
concepts).
Rather, an infinitely intelligent
mathematician should be able to derive the entire theory tree from 
these equations alone, by deriving the properties of the 
Universe that they describe, the properties of its inhabitants and their 
perceptions of the world.

The first 100 years of the quantum have provided 
powerful new technologies
and answered many questions. However, physics has raised
new questions just as important as those outstanding
at the time of Maxwell's inaugural speech;
questions regarding both quantum gravity and 
the ultimate nature of reality.
If history is anything to go by, the coming century should 
be full of exciting surprises.


\bigskip
{\bf The authors: }
MAX TEGMARK and JOHN ARCHIBALD WHEELER discussed
quantum mechanics extensively during Tegmark's three
and a half years as a postdoc at the Institute for Advanced Study
in Princeton, N.J.
Max Tegmark is now an assistant professor of physics at the 
University of Pennsylvania. Wheeler is
professor emeritus of physics at Princeton, 
where his graduate students included Richard Feynman
and Hugh Everett III (inventor of the many-worlds interpretation).
He received the 1997 Wolf Prize in physics for his work
on nuclear reactions, quantum mechanics and black holes.

The authors wish to thank
Emily Bennett and Ken Ford for help with an earlier 
manuscript on this topic, and Graham Collins, Jeff Klein, 
George Musser, Dieter Zeh and Wojciech Zurek 
for helpful suggestions.

\bigskip
\bigskip
\bigskip

{\bf Scientific American Website:} {\it www.sciam.com}
\end{document}